\documentclass[12pt, a4paper, oneside]{article}
\usepackage[top=3cm, bottom=3cm, left = 2cm, right = 2cm]{geometry} 
\usepackage{graphicx} 
\usepackage{xfrac} 
\usepackage{amsmath} 
\usepackage{amsfonts} 
\usepackage{bm}
\usepackage{comment}
\usepackage{longtable}

\usepackage{footnote}
 \usepackage{multirow}
 \usepackage{comment}

\usepackage{algorithm}
\usepackage{algorithmic}
\usepackage{mathtools} 
\usepackage{subcaption} 

\usepackage{xcolor}

\usepackage{authblk}
  \usepackage{xcolor, soul} 
\sethlcolor{lightgray}  

\definecolor{myred}{RGB}{128,22,56} 
\usepackage[colorlinks = true,
    linkcolor = myred,
    anchorcolor = myred,
    citecolor = myred,
    filecolor = myred,
    urlcolor = myred]{hyperref} 

  \usepackage{natbib}  
\title{A comparison of variable selection methods and predictive models for postoperative bowel surgery complications}

\author[1]{\"{O}zge \c{S}ahin}
\author[2]{Annemiek Kwast}
\author[3]{Annemieke Witteveen}
\author[1]{Tina Nane}

\affil[1]{\footnotesize Delft Institute of Applied Mathematics, Delft University of Technology, The Netherlands}
\affil[2]{\footnotesize Value-Based Healthcare Group, Medisch Spectrum Twente, The Netherlands}
\affil[3]{\footnotesize Biomedical Signals and Systems, Technical Medical Centre, University of Twente, The Netherlands}
\date{\today}

\begin{document} 
\maketitle

\vspace{-1.25cm}

\begin{abstract}
Accurate prediction of postoperative complications can support personalized perioperative care. However, in surgical settings, data collection is often constrained, and identifying which variables to prioritize remains an open question. We analyzed 767 elective bowel surgeries performed under an Enhanced Recovery After Surgery protocol at Medisch Spectrum Twente (Netherlands) between March 2020 and December 2023. Although hundreds of variables were available, most had substantial missingness or near-constant values and were therefore excluded. After data preprocessing, 34 perioperative predictors were selected for further analysis. Surgeries from 2020 to 2022 ($n=580$) formed the development set, and 2023 cases ($n=187$) provided temporal validation. We modeled two binary endpoints: any and serious postoperative complications (Clavien-Dindo $\ge$ IIIa). We compared weighted logistic regression, stratified random forests, and Naive Bayes under class imbalance (serious complication rate $\approx$11\%; any complication rate $\approx$35\%). Probabilistic performance was assessed using class-specific Brier scores. We advocate reporting probabilistic risk estimates to guide monitoring based on uncertainty. Random forests yielded better calibration across outcomes. Variable selection modestly improved weighted logistic regression and Naive Bayes but had minimal effect on random forests.  Despite single-center data, our findings underscore the value of careful preprocessing and ensemble methods in perioperative risk modeling.

\end{abstract}

\newpage
\section{Introduction}\label{sec:intro}
Personalized perioperative care is increasingly viewed as essential for maintaining clinical quality and the economic sustainability of healthcare systems~\citep{snyderman2012personalized}. Prognostic (risk‐prediction) models play a central role in this transition: by quantifying the probability that an individual patient will experience a specific adverse event, they can support decisions on discharge timing and resource allocation. Bowel surgery provides a good illustration: postoperative complications, such as anastomotic leakage, bleeding, or ileus, increase morbidity, prolong hospital stays, and raise healthcare costs~\citep{wen2021machine}. Even though Enhanced Recovery After Surgery (ERAS) protocols have improved patient outcomes and shortened average length of stay~\citep{Ljungqvist2017,greer2018enhanced}, clinicians still need reliable tools to distinguish high from low risk patients in order to prioritize care and optimize resources.

A range of statistical and machine learning approaches has been employed for binary risk prediction, including logistic regression, decision-tree ensembles, and generative classifiers~\citep{steyerberg2014risk,weller2018leveraging, christodoulou2019systematic}. Logistic regression remains the most widely used model in clinical research due to its interpretability and ease of implementation~\citep{boateng2019review}. However, its reliance on a linear predictor may limit its flexibility in capturing complex relationships, such as nonlinear effects, tail dependence, or interactions between continuous and discrete variables. Random forests can model these patterns more effectively but may yield poorly calibrated probabilities in sparse regions of the predictor space~\citep{leonard2022machine}. Naive Bayes classifiers, despite the strong independence assumption, can perform competitively in settings with small sample sizes and many discrete predictors~\citep{corzo2023bayesian}. They have been shown to outperform logistic regression in terms of predictive accuracy under such constraints~\citep{ng2001discriminative}.

Variable selection also remains an essential task, particularly in healthcare datasets where hundreds of perioperative predictors may be available. Selecting a relevant subset of predictors can improve model interpretability, reduce overfitting, and guide data collection towards clinically informative variables~\citep{cueto2019comparative,tian2024prediction}. Several variable selection strategies have been proposed in the literature, including wrapper methods that rely on model-based evaluation and filtering methods that use criteria such as (conditional) mutual information. In parallel, many healthcare applications suffer from class imbalance. For example, only 6.3\% of patients experience postoperative ileus following colorectal surgery~\citep{brydges2024testing}. Class imbalance can severely limit the performance of predictive models unless adjustments, such as weighting schemes or stratified sampling, are implemented~\citep{khalilia2011predicting,brydges2024testing,kuo2023artificial}.

In this study, we analyze data from 767 patients who underwent elective bowel surgery at Medisch Spectrum Twente (MST), the Netherlands, between March 2020 and December 2023. All patients were treated under the ERAS protocol~\citep{greer2018enhanced}. The raw dataset included 337 variables per patient. After preprocessing, including missing value imputation with conditional tables, we had 34 variables for analysis. We investigate two binary outcomes: the occurrence of any postoperative complication and the occurrence of a postoperative serious complication within 30 days after surgery. For model development and validation, we perform a temporal split. Surgeries performed between March 2020 and December 2022 constitute the training (in-sample) set, while those from January to December 2023 serve as the test (out-of-sample) set.  In the in-sample data, 33\% of surgeries experienced any complications, compared to 38\% in the out-of-sample data. The percentage of surgeries with serious complications is 11\% in both datasets. Hence, we perform imbalanced classification for both outcomes. 

For data exploration, we use empirical logit plots to examine the marginal risk contributions of variables to the prediction of outcomes and highlight how important predictors can vary depending on the outcome definition. For instance, we find that body mass index does not have a clear marginal impact on the outcomes, but it becomes informative when modeled conditionally alongside procedure type.

Our aim is to compare several probabilistic classification models, weighted logistic regression, random forests, and Naive Bayes, and investigate the extent to which variable selection can improve their probabilistic performance. If the latter, data collection processes can benefit from them. In particular, we analyze wrapper and filtering approaches for variable selection and assess model performance using class-specific Brier scores to account for imbalanced outcomes in addition to the area under the receiver operating characteristic curve. Random forests achieved the lowest and similar class-specific Brier scores across classes for predicting serious complications when all 34 preprocessed variables were included. Weighted logistic regression, adjusted for class imbalance, outperformed the unweighted version, while Naive Bayes showed similar predictive accuracy. Wrapper selection slightly improved the performance of weighted logistic regression but had a minimal impact on random forests.

While none of the individual modeling approaches we apply are novel, our contributions are threefold. First, we provide a systematic evaluation of variable selection strategies in a realistic perioperative setting, comparing their implications for model performance and parsimony. Second, we propose a hybrid selection strategy that combines variable exclusion with a data-driven stopping criterion based on the elbow point of the conditional mutual information curve. Third, by comparing prediction models across two outcomes, any complication versus serious complication, we show that model performance and variable importance are outcome-dependent. Our results underscore the need to tailor predictive models to specific clinical endpoints.

The remainder of the paper is structured as follows. Section~\ref{sec:methods} presents the classification models, variable selection approaches, and performance measures. Section~\ref{sec:data} details the patient cohort and provides exploratory data analyses. Section~\ref{sec:result} reports the empirical findings, and Section~\ref{sec:conc} discusses the clinical implications and potential directions for future research.

\section{Methodological framework}\label{sec:methods}

In this section, we outline a classification framework for postoperative risk assessment that we apply in Section \ref{sec:result}. In addition, we provide an overview of classification measures that can be used to assess model calibration and discrimination in imbalanced classification problems.

\subsection{Classification models and wrapper-based variable selection}\label{sec:modelingframe}
Let $\mathbf{X} = (X_1, \dots, X_d)$ be a random vector of $d$ perioperative variables taking values in $\mathbb{R}^d$, and let $Y \in \{0,1\}$ be a random variable indicating the development of a postoperative complication. The random vector may represent demographic, preoperative, and intraoperative information, and the variables can be continuous, discrete or mixed.

Suppose that we have data from $n$ patients. For each patient $i$, for $i = 1,\ldots,n$, we observe realizations $(\mathbf{x}_i, y_i)$ from $(\mathbf{X}, Y)$, where $\mathbf{x}_i = (x_{i,1}, \dots, x_{i,d})$. We aim to construct a classifier that accurately predicts $y_i$ given new observations $\mathbf{x}_i$. Further, our interest is also to get a probability estimate denoted by 
$
\hat{p}(y_i \mid\mathbf{x}_i) = \widehat{\Pr}(Y = y_i \mid \mathbf{X} = \mathbf{x}_i).
$
Such probability estimates are more informative than binary outcomes for patient risks. For example, one can consider that an estimate of 0.95 indicates a higher risk than an estimated probability of 0.55 for developing complications. Still, such probability estimates can be thresholded (typically at 0.50 for binary outcomes) to assign a predicted class: $  f(\mathbf{x}_i)= 
\begin{cases}
1, & \text{if } \hat{p}(\mathbf{x}_i)\geq 0.50,\\
0, & \text{otherwise}.
\end{cases}
$

Hence, the goal is to identify a function $f: \mathbb{R}^d\to\{0,1\}$ that classifies a patient's postoperative complications as accurately as possible. 

Moreover, it is likely that some variables are redundant or do not have any relationship with the outcome, especially if we have a high number of perioperative variables. For instance, one could have multiple variables indicating the patient's current health status, such as the American Society of Anesthesiologists Physical Status (ASA) or the World Health Organization (WHO) scores, and keeping one would be enough. Thus, we must decide which variables to keep. To address this, wrapper-based variable selection methods can be used. 

Formally, let $S \subseteq \{1,\dots,d\}$ represent a subset of the $d$ variables, and let $L(S)$ denote a performance metric (e.g., classification accuracy or weighted log-likelihood) obtained by training and testing the model on variables indexed by $S$. Then, we analyze 
$
S^{*} \;=\;\underset{S}{\mathrm{argmax}}\;L(S).
$
Since searching all subsets is typically infeasible, one can iteratively expand or reduce $S$, and each time, one assesses $L(\cdot)$ by training and testing the classifier on a portion of the data. Then, variables are kept or removed based on their contribution to $L(\cdot)$. Hence, the final selected subset $S^{*}$ is directly tied to the predictive accuracy of the chosen model.

\subsubsection*{Classification evaluation metrics}\label{sec:measures}
To compare the models' classification performance regarding their discrimination and calibration, we will report the area under the receiver operating characteristic curve (AUC) for the former. In addition, we will work with the Brier score  \citep{brier1950verification} for the former and the latter. While AUC is not a probabilistic score, the Brier score allows us to consider the probabilistic comparison of the models.

The Brier score for binary classification is defined as  $BS =\frac{1}{n} \sum_{i=1}^{n}(j-\hat{p}(y_i = j \mid\mathbf{x}_i) )^{2}, \quad j \in \{0,1\}.$ To consider the imbalanced classification, we will also consider the Brier score in both patient populations (classes), namely 

\begin{equation}
BS_j = \frac{1}{n_j} \sum_{y_i=j}(y_{i}-\hat{p}(y_i=j \mid\mathbf{x}_i) )^{2},
    \label{eq:Brier_strat} \qquad \textrm{for} \qquad j \in \{0,1\},
\end{equation}
for $n_j=\sum_{i=1}^n \mathbf{1}_{\{y_i=j\}}$. Here, $j\in\{0,1\}$ can refer to developing (1) and not developing (0) any complication or developing ($1$) and not developing ($0$) a serious complication after surgeries. 

\subsubsection*{(Weighted) Logistic regression}\label{sec:logistic}

Logistic regression is arguably the most popular method to model binary random variables via the relationship to multiple predictor variables. It estimates the probability. 

\begin{equation}
    \label{eq:logistic}
    P(Y_i = 1 \mid \mathbf{X} = \mathbf{x}_i)  = \frac{1}{1+\exp(-\boldsymbol{\beta}^T \mathbf{x}_i)}=\pi_i,
\end{equation}
for $i=1,\ldots,n$, by considering a maximum likelihood estimate $\boldsymbol{\hat{\beta}}$  of the coefficient vector $\boldsymbol{\beta}$  of length $d+1$. The resulting $\exp(\hat{\beta}_j)$'s, for $j=1,\ldots,d$ give the change in the ratio of odds for a unit increase in $X_j$, while the other predictors remain unchanged. Similarly, $\exp(\hat{\beta}_0)$ gives the ratio of odds when all $X_j$'s are zero.   

The log-likelihood is given by, for a sample of size $n$, \[
l(\boldsymbol{\beta}; y, \mathbf{x}) = \sum_{i| y_i=1} \ln(\pi_i) + \sum_{i| y_i=0} \ln(1-\pi_i) = -\sum_{i=1}^n \ln(1+\exp((1-2y_i)\boldsymbol{\beta}^T \mathbf{x}_i)).
\]

To account for the imbalanced classes, we consider weighted logistic regression \citep{king2001logistic}, for which the weighted log-likelihood function is given by

\[
l_w(\boldsymbol{\beta}; y, \mathbf{x}) = w_1\sum_{i| y_i=1} \ln(\pi_i) + w_0\sum_{i| y_i=0} \ln(1-\pi_i)= -w_i\sum_{i=1}^n \ln(1+\exp((1-2y_i)\boldsymbol{\beta}^T \mathbf{x}_i)),
\]
where $w_0+w_1=1$, and $w_i=w_1y_i+w_0(1-y_i)$. Weights $w_0$ and $w_1$ are chosen to increase the ``importance'' of the imbalanced class observations within the log-likelihood. The \texttt{R} package \texttt{stats} \citep{stats} has built-in functions to accommodate for weighted logistic regression. We choose the weights to be equal to the inverse of the frequencies of the two classes (e.g., any complication vs non-complication) in Section \ref{sec:result}. 

For wrapper-based variable selection, we also applied stepwise variable–selection procedures for (weighted) logistic regression models (forward, backward, and both). Starting from the null model (forward), the full model (backward), or whichever of the two leads to the larger improvement at a given step (both), predictors were added or removed one at a time according to the change in the Akaike Information Criterion \citep{akaike1987factor} in the training set. The subset that minimized this criterion was then refitted, with the previously described class weights, to the entire training data before the out-of-sample assessment. The \texttt{R} package \texttt{MASS} \citep{MASS} has built-in functions to accommodate such variable selection.

We remark that throughout the (weighted) logistic regression analyses, we specify all ordinal and nominal predictors as factor variables. This specification ensures that the model estimates separate coefficients for each category without imposing unjustified linearity across their levels.

\subsubsection*{Random forests}\label{sec:decisiontree}
Random forests are ensemble methods that extend the idea of decision trees. In a decision tree, the $d$-dimensional variable space is recursively partitioned via binary splits into terminal nodes (or leaves). Terminal nodes, then, are the unsplit regions used for final prediction.  Suppose node $t$  contains $n_t$ observations, with $n_{t,0}$ observations from class 0 and $ n_{t,1}$ from class 1. We define the class proportions at node $t$  as $p_{t,0} = \frac{n_{t,0}}{n_t} \quad \text{and} \quad p_{t,1} = \frac{n_{t,1}}{n_t}.$ At each node $t$ of a decision tree, the algorithm selects the variable and splitting point that provide the largest decrease in node impurity. A common measure of impurity is the Gini index \citep{breiman2017classification}, defined for node $t$ as 
$G(t) = 1 - p_{t,0}^2 - p_{t,1}^2.$
When node $t$  is split into two nodes $t_L$ and $t_R$, with $n_{t_L}$ and $n_{t_R}$ observations respectively, the weighted impurity after the split is given by
$
I_{\text{split}} = \frac{n_{t_L}}{n_t} \, G(t_L) + \frac{n_{t_R}}{n_t} \, G(t_R).
$
Then, the split is chosen to maximize the decrease in impurity, namely, to maximize
$\Delta G = G(t) - I_{\text{split}}$. A lower Gini index indicates that the observations in the node predominantly belong to one class. Further, to control tree complexity and reduce overfitting, tree-growing criteria, such as setting a maximum number of terminal nodes and a minimum number of observations per terminal node, are typically used. 

In Section~\ref{sec:result}, we implement random forests using the \texttt{randomForest} package in \texttt{R} \citep{randomForest}, setting the minimum terminal node size to 50 and applying stratified sampling to address class imbalance \citep{aoyama1954study}. For wrapper-based variable selection, we use 10-fold cross-validation in the in-sample data to evaluate random forests trained on subsets of 5, 10, 15, and 20 variables. The subset yielding the highest validation accuracy (on the portion of the in-sample data) is selected.  However, one could extend this analysis to any subset size as long as computational resources are available. We implement these ideas using the \texttt{R} package \texttt{caret} \citep{caret}.

\subsubsection*{Naive Bayes with kernel density estimation}\label{sec:nbkde}
Another way to obtain $\hat{p}(\mathbf{x}_i)$ is via discriminant analysis methods based on Bayes' rule. For each class $j \in \{0,1\}$, we estimate
\begin{equation}
\label{eq:bayesrule}
\widehat{\Pr}(Y_i = j \mid \mathbf{X}_i = \mathbf{x}_i) 
= 
\frac{\hat{\pi}_j \cdot \widehat{f}_{\mathbf{X}\mid J}(\mathbf{x}_i \mid j)}{\sum_{j'} \hat{\pi}_{j'} \cdot \widehat{f}_{\mathbf{X}\mid J}(\mathbf{x}_i \mid j')},
\end{equation}
where $\widehat{f}_{\mathbf{X}\mid J}(\cdot \mid j)$ is the estimated conditional density for the class $j$, and $\hat{\pi}_j$ is the prior probability of class $j$. The latter is often taken as equal or proportional to observed class frequencies. If some perioperative variables are continuous and others are ordinal, $\widehat{f}_{\mathbf{X}\mid J}(. \mid j)$ can be modeled by vine copulas to account for dependence among variables \citep{csahin2024vine}. Alternatively, one may assume conditional independence across variables, which leads to a Naive Bayes approach:
\begin{equation}
\label{eq:NBproduct}
\widehat{f}_{\mathbf{X}\mid J}(\mathbf{x}_i \mid j) 
= 
\prod_{p=1}^{d}\,\widehat{f}_{X_{p}}\bigl(x_{i,p} \mid j\bigr),
\end{equation}
where $\widehat{f}_{X_{p}}(\cdot \mid j)$ is the $p$th marginal density estimate fitted for class $j$. When these marginal densities are fitted with kernel density estimation, the method is called Naive Bayes with kernel density estimation \citep{csahin2024vine}. Implementations of both the vine copula approach and Naive Bayes are available in the \texttt{R} package \texttt{vineclass} \citep{vineclass}. Finally, because each class' density is considered and weighted by its prior in Equation \eqref{eq:bayesrule}, Naive Bayes with kernel density estimation inherently adjusts for imbalanced classes. 

As a wrapper-based variable selection, we split the training data into a learning set (75\%) and a validation set (25\%). Next,  we optimize a criterion \(L(\cdot)\) defined as the average Brier score calculated separately for each class (Equation \eqref{eq:Brier_strat}) . In our case study in Section \ref{sec:result}, it is the average Brier scores of the (non-serious) non-complications and (serious) complication classes. To be more concrete, we identify the two variables that give the best $L(.)$ on the learning set. We then iteratively add one variable at a time and choose the variable whose inclusion results in the highest improvement in $L(.)$ on the learning set. We continue this process until adding more variables fails to improve $L(.)$  on the validation set. Then, the already chosen variables, until stopping, give the final subset of variables.

\subsection{Filtering based variable-selection}\label{sec:filter}
We also consider model-independent variable selection techniques, known as filtering methods. Unlike wrapper approaches, which iteratively train classification models, filtering approaches rely on statistical or information-theoretic criteria and are thus computationally efficient. Among various filtering methods benchmarked in \cite{bommert2020benchmark}, no single method consistently outperformed across datasets. Motivated by their flexibility, we use mutual information (MI) and conditional mutual information (CMI) to quantify the informativeness of a variable for predicting the outcome. As most perioperative variables are discrete or discretized in clinical practice, we adopt discrete formulations. Let $X$ and $Y$ be discrete random variables representing a variable and a class label, respectively. Let $f_X(x)$ and $f_Y(y)$ denote the probability mass functions (pmfs) for $X$ and $Y$, and let $f_{XY}(x,y)$ be their joint pmf. Then, the MI between $X$ and $Y$ is
\begin{equation}
\label{eq:MI-discrete}
I(X, Y) 
= \sum_{x,y} f_{XY}(x, y)\,\log\!\Bigl(\frac{f_{XY}(x, y)}{f_X(x)\,f_Y(y)}\Bigr).
\end{equation}
MI quantifies how much knowledge of $X$ reduces uncertainty about $Y$, and a higher $I(X, Y)$ implies that the variable $X$ is more informative for predicting $Y$. To evaluate conditional informativeness, the CMI of $X$ and $Y$ given $\mathbf{Z}$ is then defined as
\begin{equation}
\label{eq:CMI-discrete}
I(X, Y \mid \mathbf{Z})
= \sum_{x,y,\mathbf{z}}\; f_{XYZ}(x, y, \mathbf{z}) \,\log\!\Bigl(
    \frac{f_{XY\mid Z}(x,y\mid \mathbf{z})}{f_{X\mid Z}(x\mid \mathbf{z})\,f_{Y\mid Z}(y\mid \mathbf{z})}
\Bigr).
\end{equation}
By conditioning on $\mathbf{Z}$, $I(X,Y\mid \mathbf{Z})$ shows how much additional information $X$ provides on $Y$ beyond what is already modeled by $\mathbf{Z}$.  

In Section~\ref{sec:data}, we adopt a data-driven stopping rule: starting from three informative variables (one via MI, two via CMI), we compute the CMI for each candidate conditional on this set, and apply a knee-elbow analysis to identify a subset. More details are provided in Appendix~\ref{sec:app-CMI}.

\section{Data}\label{sec:data}

\subsection{Data overview}\label{sec:dataoverview}
We analyzed patients who underwent elective bowel surgery at Medisch Spectrum Twente (MST), the Netherlands, between March 2020 and December 2023. All patients were treated according to the ERAS protocol. The raw data included 767 surgeries (observations) and 337 recorded variables.  

Two binary outcomes were available: (i) occurrence of any postoperative complication and (ii) occurrence of a serious complication within 30 days of surgery. Complication severity was coded using the Clavien-Dindo (CD) classification \citep{clavien1992proposed}. Throughout, we label a complication as serious if CD~$\ge$~IIIa; otherwise, complications (including none) are considered non-serious. For illustration, an intraperitoneal abscess (often requiring surgical drainage) is classified as serious, whereas an ileus managed conservatively with medication is not. Even though the database has specific complication types, counts were too small to support subtype-specific modelling (e.g., postoperative ileus occurred in only 23 patients).

For model development and validation, we split the data temporally: surgeries from March 2020 to December 2022 formed the training (in-sample) set (518 non-serious, 62 serious; 390 non-complication, 190 any complication), and surgeries from January 2023 to December 2023 formed the test (out-of-sample) set (168 non-serious, 19 serious; 116 non-complication, 71 any complication). This design allows us to assess temporal generalization and model validation within the same clinical setting. Given that the number of any or serious complications is much less than non-complication and non-serious, we have an imbalanced classification problem for both outcomes.

This study received ethical approval from the niet-WMO adviescommissie MST  (approval K24-41) on Month 12, 2024. This is an IRB-approved retrospective study; all patient information was de-identified, and patient consent was not required. We used the TRIPODreporting guidelines \citep{collins2015transparent}.

\subsection{Data cleaning and preprocessing}
We conducted variable-level data cleaning to ensure quality and relevance. First, we excluded over 170 variables indicating specific postoperative complications or CD scores since they are outcomes rather than predictors. Similarly, over 50 postoperative variables (e.g., nausea severity, daily weight changes) were removed due to their timing relative to complications. Variables with more than 50\% missingness (e.g., glucose levels, preoperative weight changes, intraoperative temperature) or near-constant values (e.g., preoperative education) were also excluded.

To address potential misclassification in outcomes, we manually reviewed 16 patients with prolonged stays and no recorded complications. For 11 of them, complications with CD = 0 or 1 were identified and used to update their status.

Missing values were imputed using conditional tables informed by clinical context. Encoding was applied where needed; for instance, patients who quit smoking just before surgery were categorized as smokers. Full imputation and encoding details are given in Table~\ref{tab:app-imputation} in Appendix~\ref{sec:app-datapreproc}.

After preprocessing, the final dataset had 34 perioperative variables and two binary outcomes for 767 patients. Table~\ref{tab:summaryin} summarizes the in-sample data used for model development; model validation uses the out-of-sample data summarized in Table~\ref{tab:summaryout}.

\subsection{Exploratory data analysis}\label{sec:EDA}

\begin{figure}[ht!]
    \centering
    \includegraphics[width=1.0\textwidth]{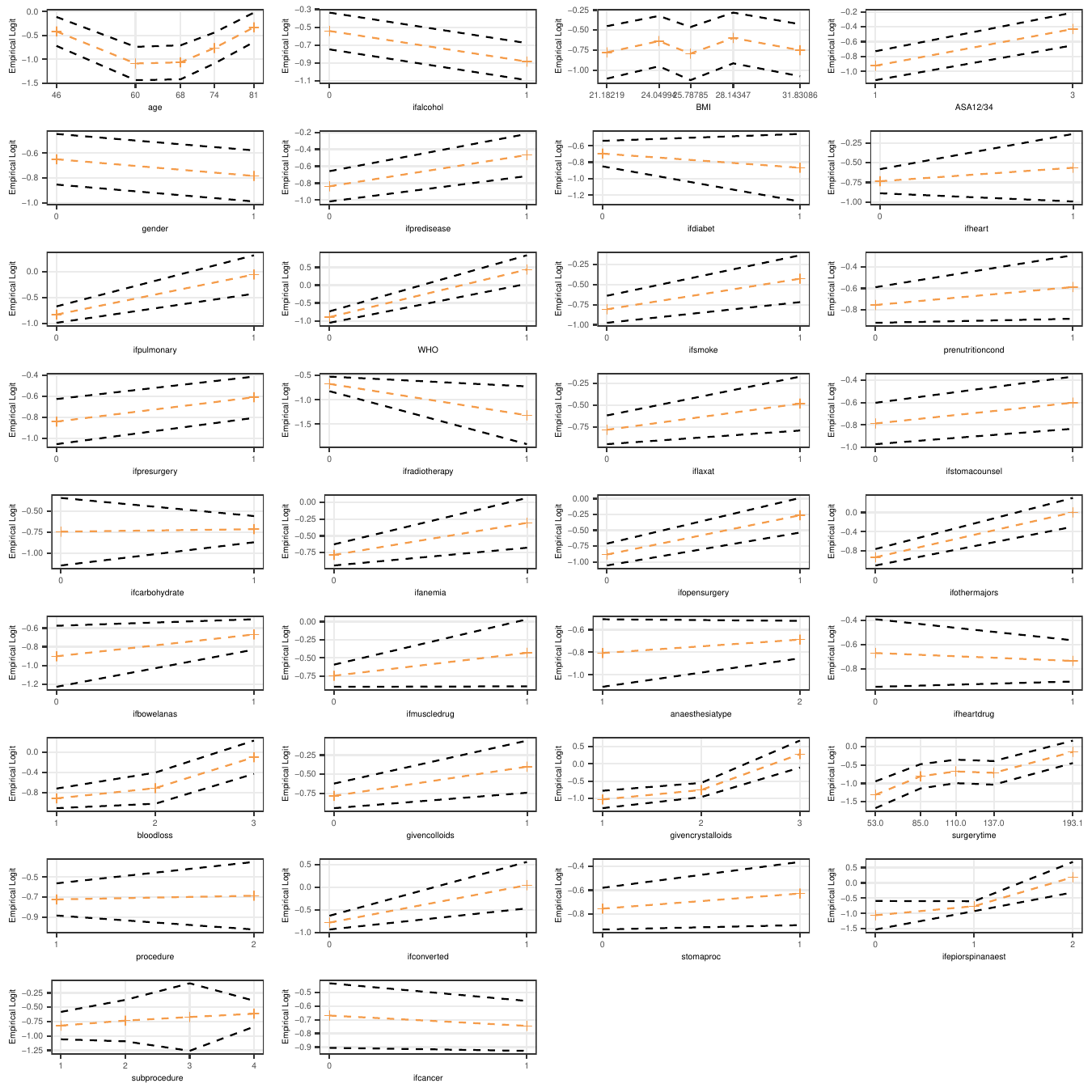}
    \caption{Empirical logit plots of variables with confidence levels for any complication likelihood across categories. Categories are plotted along the x-axis with corresponding logits on the y-axis. For interpretability, continuous variables are discretized into quantile-based intervals, specifically at the 10th, 30th, 50th, 70th, and 90th percentiles. The dashed orange line represents the empirical logit values, while the black dashed lines indicate the 90\% confidence limits for these estimates. Points correspond to the midpoints of the discretized intervals.}
    \label{fig:emploganycomp}
\end{figure}

\begin{figure}[ht!]
    \centering
    \includegraphics[width=1.0\textwidth]{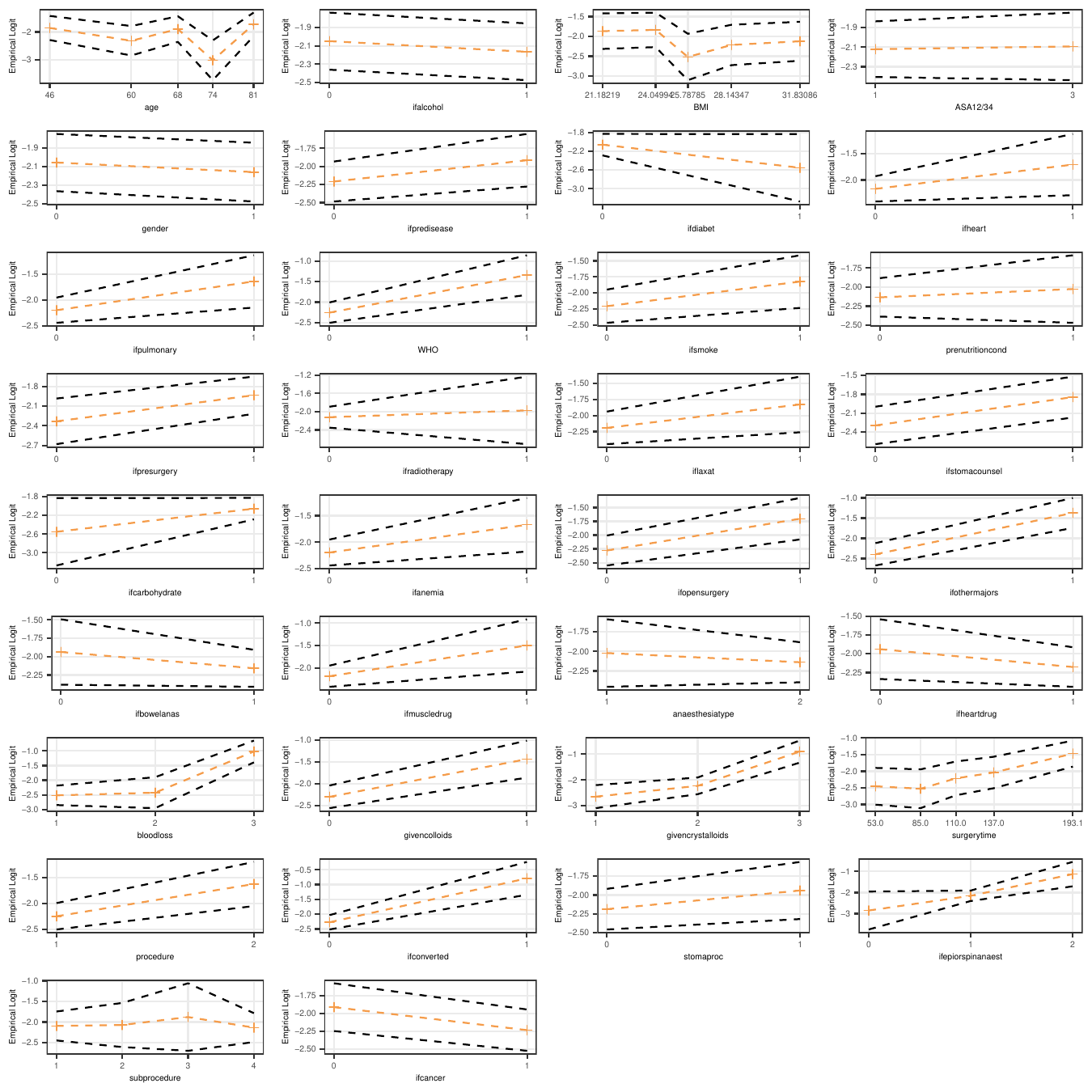}
    \caption{Empirical logit plots of variables with confidence levels for serious complication likelihood across categories. Categories are plotted along the x-axis with corresponding logits on the y-axis. Continuous variables are discretized into quantile-based intervals, specifically at the 10th, 30th, 50th, 70th, and 90th percentiles for interpretability. The dashed orange line represents the empirical logit values, while the black dashed lines indicate the 90\% confidence limits for these estimates. Points correspond to the midpoints of the discretized intervals.}
    \label{fig:emplogseriouscomp}
\end{figure}

Table \ref{tab:summaryin} provides an overview of the variables, including their ranges and estimated mean values across the classes of patients who developed or did not develop any complications and serious complications. To better explore the marginal impact of each variable on the likelihood of developing (serious) complications, we conducted an empirical logit analysis presented in Figures \ref{fig:emploganycomp} and \ref{fig:emplogseriouscomp}.

\begin{table}[ht!]
\centering
\tiny

\resizebox{\textwidth}{!}{%
\begin{tabular}{llllll}
\hline
\textbf{Variable} & 
  \textbf{$j=0$} & 
  \textbf{$j=1$ } &
  \textbf{$j=0_s$} &
  \textbf{$j=1_s$} &
  \textbf{Description} \\
\hline

\texttt{age} 
 & 65.2 [20--91] 
 & 65.5  [24--92] 
 & 65.4 [20--92] 
 & 64.3  [24--88] 
 & Age of the patient \\

\texttt{ifalcohol}
 & 216 (55.3\%) 
 & 89 (46.8\%) 
 & 274 (52.9\%) 
 & 31 (50.0\%)
 & Alcohol consumption indicator \\

\texttt{BMI}
 & 26.3  [14.03--48.33]
 & 26.4  [14.47--40.06]
 & 26.4  [14.03--48.33]
 & 25.9  [14.47--39.79]
 & Body mass index \\

\hline
\texttt{ASA12/34}
 & \begin{tabular}[c]{@{}l@{}}
    1: 250 (64.1\%)\\
    3: 140 (35.9\%)
   \end{tabular}
 & \begin{tabular}[c]{@{}l@{}}
    1: 99 (52.1\%)\\
    3: 91 (47.9\%)
   \end{tabular}
 & \begin{tabular}[c]{@{}l@{}}
    1: 312 (60.2\%)\\
    3: 206 (39.8\%)
   \end{tabular}
 & \begin{tabular}[c]{@{}l@{}}
    1: 37 (59.7\%)\\
    3: 25 (40.3\%)
   \end{tabular}
 & ASA score \{1,2\} vs. \{3,4\} \\

\hline
\texttt{gender}
 & 202 (51.8\%) 
 & 92 (48.4\%) 
 & 264 (51.0\%) 
 & 30 (48.3\%) 
 & Male (1) or female (0) \\

\texttt{ifpredisease}
 & 112 (28.8\%)
 & 70 (36.8\%)
 & 159 (30.7\%)
 & 23 (37.1\%)
 & Pre-existing disease indicator \\

\texttt{ifdiabet}
 & 53 (13.6\%)
 & 22 (11.6\%)
 & 70 (13.5\%)
 & 5 (8.6\%)
 & Diabetes indicator \\

\texttt{ifheart}
 & 39 (10.0\%)
 & 22 (11.6\%)
 & 52 (10.0\%)
 & 9 (14.5\%)
 & Heart condition indicator \\

\texttt{ifpulmonary}
 & 39 (10.0\%)
 & 37 (19.5\%)
 & 64 (12.3\%)
 & 12 (19.4\%)
 & Pulmonary condition indicator \\

\texttt{WHO}
 & 27 (6.9\%)
 & 42 (22.1\%)
 & 55 (10.6\%)
 & 14 (22.6\%)
 & WHO performance score \\

\texttt{ifsmoke}
 & 80 (20.5\%)
 & 52 (27.4\%)
 & 114 (22.0\%)
 & 18 (29.0\%)
 & Smoking status \\

\texttt{prenutritioncond}
 & 85 (21.8\%)
 & 47 (24.7\%)
 & 117 (22.6\%)
 & 15 (24.1\%)
 & Pre-surgical nutrition condition \\

\texttt{ifpresurgery}
 & 197 (50.5\%)
 & 107 (56.3\%)
 & 266 (51.4\%)
 & 38 (61.2\%)
 & Indicator of previous surgeries \\

\texttt{ifradiotherapy}
 & 35 (9.0\%)
 & 9 (4.7\%)
 & 39 (7.5\%)
 & 5 (8.1\%)
 & Radiotherapy indicator \\

\texttt{iflaxat}
 & 73 (18.7\%)
 & 45 (23.7\%)
 & 102 (19.7\%)
 & 16 (25.8\%)
 & Oral laxatives indicator \\

\texttt{ifstomacounsel}
 & 139 (35.6\%)
 & 76 (40.0\%)
 & 186 (35.9\%)
 & 29 (46.7\%)
 & Stoma counseling indicator \\

\texttt{ifcarbohydrate}
 & 339 (86.9\%)
 & 166 (87.4\%)
 & 448 (86.5\%)
 & 57 (92.0\%)
 & Carbohydrate intake indicator \\

\texttt{ifanemia}
 & 45 (11.5\%)
 & 33 (17.4\%)
 & 66 (12.7\%)
 & 12 (19.4\%)
 & Anemia indicator \\

\texttt{ifopensurgery}
 & 82 (21.0\%)
 & 63 (33.1\%)
 & 123 (23.8\%)
 & 22 (35.5\%)
 & Open surgery indicator \\

\texttt{ifothermajors}
 & 60 (15.3\%)
 & 60 (31.6\%)
 & 96 (18.5\%)
 & 24 (38.7\%)
 & Multiple major procedures indicator \\

\texttt{ifbowelanas}
 & 303 (77.7\%)
 & 155 (81.6\%)
 & 411 (79.3\%)
 & 47 (75.8\%)
 & Bowel anastomosis indicator \\

\texttt{ifmuscledrug}
 & 31 (8.0\%)
 & 20 (10.5\%)
 & 42 (8.1\%)
 & 9 (14.5\%)
 & Muscle relaxants indicator \\

\hline
\texttt{anaesthesiatype}
 & \begin{tabular}[c]{@{}l@{}}
    1: 97 (24.9\%)\\
    2: 293 (75.1\%)
   \end{tabular}
 & \begin{tabular}[c]{@{}l@{}}
    1: 43 (22.6\%)\\
    2: 147 (77.4\%)
   \end{tabular}
 & \begin{tabular}[c]{@{}l@{}}
    1: 124 (23.9\%)\\
    2: 394 (76.1\%)
   \end{tabular}
 & \begin{tabular}[c]{@{}l@{}}
    1: 16 (25.8\%)\\
    2: 46 (74.2\%)
   \end{tabular} 
 & Inhalational (1) or TIVA (2) \\

\hline
\texttt{ifheartdrug}
 & 288 (73.9\%)
 & 138 (72.6\%)
 & 383 (73.9\%)
 & 23 (69.3\%)
 & Heart medication indicator \\

 \hline
\texttt{bloodloss}
 & \begin{tabular}[l]{@{}l@{}}
   1: 252 (64.6\%)\\
   2: 86 (22.1\%)\\
   3: 52 (13.3\%)
   \end{tabular}
 & \begin{tabular}[l]{@{}l@{}}
   1: 101 (53.2\%)\\
   2: 42 (22.1\%)\\
   3: 47 (24.7\%)
   \end{tabular}
 & \begin{tabular}[l]{@{}l@{}}
   1: 327 (63.1\%)\\
   2: 118 (22.8\%)\\
   3: 73 (14.1\%)
   \end{tabular}
 & \begin{tabular}[l]{@{}l@{}}
   1: 26 (41.9\%)\\
   2: 10 (16.1\%)\\
   3: 26 (41.9\%)
   \end{tabular}
 & \begin{tabular}[l]{@{}l@{}}
   1:  No blood loss \\
   2: $\leq 100$ mL blood loss\\
   3: $>100$ mL blood loss
   \end{tabular} 
 \\

 \hline
\texttt{ifgivencolloids}
 & 57 (14.6\%)
 & 38 (20.0\%)
 & 77 (14.9\%)
 & 18 (29.3\%)
 & Colloids given indicator \\

 \hline
\texttt{givencrystalloids}
 & \begin{tabular}[l]{@{}l@{}}
   1: 161 (41.3\%)\\
   2: 199 (51.0\%)\\
   3: 30  (7.7\%)
   \end{tabular}
 & \begin{tabular}[l]{@{}l@{}}
   1: 57 (30.0\%)\\
   2: 93 (48.9\%)\\
   3: 40 (21.1\%)
   \end{tabular}
 & \begin{tabular}[l]{@{}l@{}}
   1: 204 (39.4\%)\\
   2: 264 (51.0\%)\\
   3: 50  (9.6\%)
   \end{tabular}
 & \begin{tabular}[l]{@{}l@{}}
   1: 14 (22.6\%)\\
   2: 28 (45.2\%)\\
   3: 20 (32.3\%)
   \end{tabular}
 & \begin{tabular}[l]{@{}l@{}}
   1: $< 1000$ mL crystalloids\\
   2: $ \geq1000 \text{ and } < 2000$ mL crystalloids\\
   3: $\geq2000$ mL crystalloids
   \end{tabular} 
 \\

 \hline

\texttt{surgerytime}
 & 110.4 [15--475]
 & 134.2 [18--383]
 & 114.6  [15--475]
 & 148.1  [32--383]
 & Duration of surgery (minutes) \\
\hline
\texttt{procedure}
 & \begin{tabular}[c]{@{}l@{}}
    1: 320 (82.1\%)\\
    2: 70 (17.9\%)
   \end{tabular}
 & \begin{tabular}[c]{@{}l@{}}
    1: 155 (81.6\%)\\
    2: 35 (18.4\%)
   \end{tabular}
 & \begin{tabular}[c]{@{}l@{}}
    1: 430 (83.0\%)\\
    2: 88 (17.0\%)
   \end{tabular}
 & \begin{tabular}[c]{@{}l@{}}
    1: 45 (72.6\%)\\
    2: 17 (27.4\%)
   \end{tabular}
 & Colonic\&small bowel(1), rectal(2) \\
\hline
\texttt{ifconverted}
 & 19 (4.9\%)
 & 20 (10.5\%)
 & 27 (5.2\%)
 & 12 (19.0\%)
 & Conversion  to open \\

\texttt{stomaproc}
 & 111 (28.5\%)
 & 59 (31.1\%)
 & 149 (28.8\%)
 & 21 (33.9\%)
 & Stoma procedure indicator \\

 \hline
\texttt{ifepiorspinanaest}
 & \begin{tabular}[c]{@{}l@{}}
    0: 47 (12.1\%)\\
    1: 324 (83.1\%)\\
    2: 19 (4.9\%)
   \end{tabular}
 & \begin{tabular}[c]{@{}l@{}}
    0: 16 (8.4\%)\\
    1: 151 (79.5\%)\\
    2: 23 (12.1\%)
   \end{tabular}
 & \begin{tabular}[c]{@{}l@{}}
    0: 60 (11.6\%)\\
    1: 426 (82.2\%)\\
    2: 32 (6.2\%)
   \end{tabular}
 & \begin{tabular}[c]{@{}l@{}}
    0: 3 (4.8\%)\\
    1: 49 (79.0\%)\\
    2: 10 (16.1\%)
   \end{tabular} 
 &  \begin{tabular}[c]{@{}l@{}}
    No epidural anesthesia (0)\\
    Thoracic or lumbar epidural anesthesia (1)  \\
  Spinal anesthesia (2) \end{tabular} \\
 \hline

\texttt{subprocedure}
 & \begin{tabular}[l]{@{}l@{}}
 1: 155 (39.7\%)\\
 2: 63 (16.2\%)\\
 3: 22 (5.6\%)\\
 4: 150 (38.5\%)
 \end{tabular}
 & \begin{tabular}[l]{@{}l@{}}
 1: 68 (35.8\%)\\
 2: 30 (15.8\%)\\
 3: 11 (5.8\%)\\
 4: 81 (42.6\%)
 \end{tabular}
 & \begin{tabular}[l]{@{}l@{}}
 1: 199 (38.4\%)\\
 2: 83 (16.0\%)\\
 3: 29 (5.6\%)\\
 4: 207 (40.0\%)
 \end{tabular}
 & \begin{tabular}[l]{@{}l@{}}
 1: 24 (38.7\%)\\
 2: 10 (16.1\%)\\
 3: 4 (6.5\%)\\
 4: 24 (38.7\%)
 \end{tabular}
 &   \begin{tabular}[l]{@{}l@{}}
 Sub-type of procedure\\
(see Appendix \ref{sec:app-datapreproc}) \\
 \\

 \end{tabular}  \\  \hline

\texttt{ifcancer}
 & 249 (63.9\%)
 & 118 (62.1\%)
 & 332 (64.1\%)
 & 35 (56.5\%)
 & Cancer indicator \\

\hline
\end{tabular}
}
\caption{Summary of variables in not developing (390 observations, $j=0$) and developing any complications (190 observations, $j=1$), as well as in not developing (518 observations, $j=0_s$) and developing serious complications classes (62 observations, $j=1_s$) in in-sample data. For each variable, it provides the distribution or summary measure (range and mean) based on the variable type, along with a brief description of the variable. For binary variables taking values of 0 and 1, the summaries for the value of 1 are reported.}
\label{tab:summaryin}
\end{table} 

Figures~\ref{fig:emploganycomp} and~\ref{fig:emplogseriouscomp} show that \texttt{age} had a nonlinear relationship with complication risks. Poorer health indicators, such as higher \texttt{ASA12/34} and \texttt{WHO} scores, were associated with a higher likelihood of any complications, even though only \texttt{WHO} was linked to serious complications. Pulmonary conditions increased risk, potentially due to either respiratory vulnerability or collinearity. Other patient characteristics, including \texttt{BMI}, alcohol use, gender, comorbidities, and nutritional status, showed no clear marginal effects.

Most preoperative variables, including radiotherapy, prior surgery, or prehabilitation interventions, showed no marginal associations, while open surgeries and multiple major procedures were linked to higher risk, suggesting potential benefit from minimally invasive techniques.

Among intraoperative factors, rectal (vs.\ colonic) procedures, high colloid use, and conversions from minimally invasive to open surgery were associated with increased risk of serious complications. High blood loss, high crystalloid volumes, and longer surgery durations (above approximately 138 minutes) were also associated with higher complication risk. Epidural anesthesia, especially thoracic or lumbar, showed increase risks, possibly reflecting case complexity. However,  we note the small sample size (14 observations) that might cause sampling variability.

\begin{figure}[ht!]
    \centering
    \includegraphics[width=.8\textwidth]{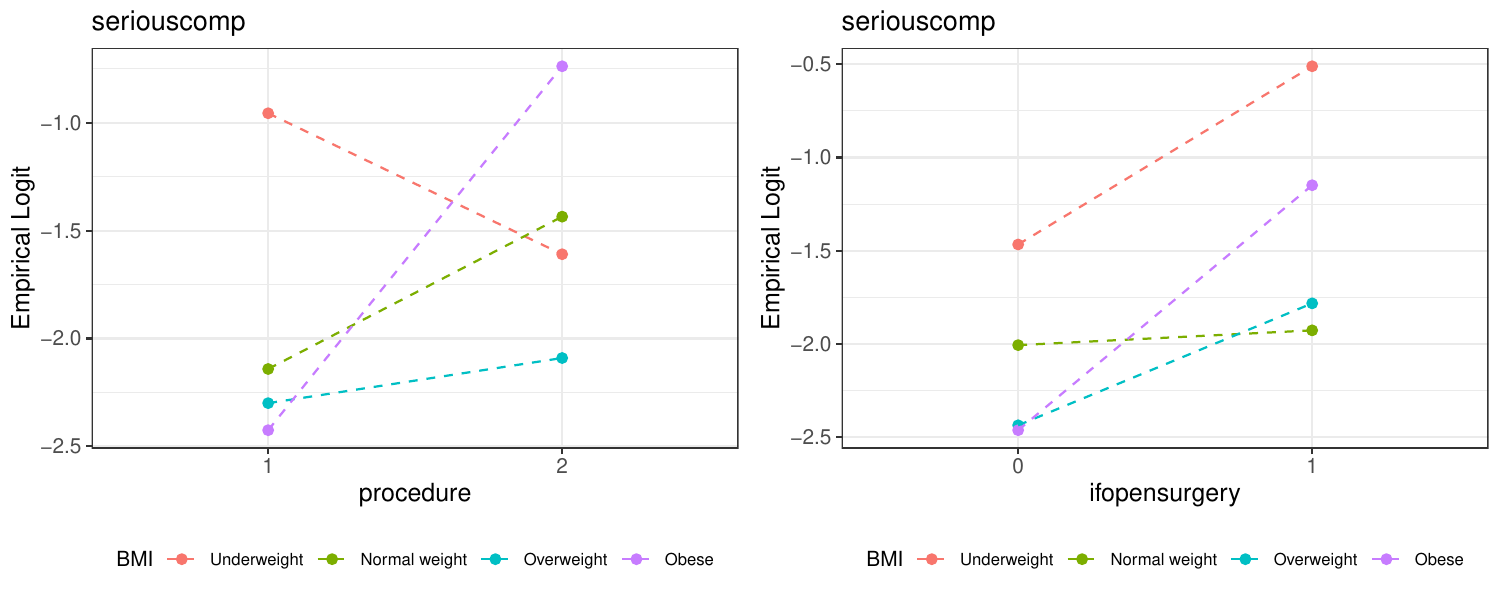}
    \caption{Empirical logit plots showing the interaction between \texttt{BMI} and \texttt{procedure} (left, 1: colonic \& small bowel, 2: rectal) and between \texttt{BMI} and \texttt{ifopensurgery} (right, 0: not-open surgery, 1: open surgery) on the likelihood of developing serious complications. The x-axis represents binary variables, while dashed lines connect the empirical logits across discretized \texttt{BMI} levels. \texttt{BMI} is grouped into four categories: "Underweight" (until 18.5), "Normal weight" (18.5 - 24.9), "Overweight" (25.0 - 29.9), and "Obese" (more than 29.9).}
    \label{fig:emploginteraction}
\end{figure}

To explore potential interaction effects between variables, we constructed empirical logit interaction plots. Continuous variables were discretized into quantiles or predefined intervals. For each pair of variable categories, we plotted the empirical logit of serious complication risk to reveal joint effects beyond univariate associations. Figure~\ref{fig:emploginteraction} illustrates interactions between \texttt{BMI} and \texttt{procedure}, and between  \texttt{ifopensurgery} and \texttt{BMI}, on serious complications. Even though \texttt{BMI} showed no clear marginal effect, its impact varied by context: obese patients undergoing rectal surgery had a higher risk than those with colonic procedures, reflecting technical challenges associated with obesity~\citep{merkow2009effect, alizadeh2016body}. Similarly, open surgery increases risks at both extremes of \texttt{BMI}, potentially due to surgical access difficulties~\citep{kassahun2022impact}. For underweight patients, the observed pattern may reflect sampling variability due to limited data (14 observations). These findings underscore the value of modeling first-order interactions in logistic regression, but random forests and Naive Bayes inherently capture such dependencies.

\section{Classification and variable selection results}\label{sec:result}

\subsection{Identification and comparison of variables for complication risks}\label{sec:varselresult}
In this section, we aim to identify the main risk factors among the considered 34 perioperative variables for developing any complications (\texttt{anycomp}) and serious complications (\texttt{seriouscomp}), when following bowel surgeries, with the data explored in Section \ref{sec:data}. As outlined in Section \ref{sec:modelingframe}, we apply filtering and wrapper approaches across three modeling frameworks. Moreover, Sections \ref{sec:predanycompresult} and \ref{sec:predseriouscompresult} also consider models that use all 34 variables for predictive modeling. 

\begin{figure}[ht!]
    \centering
    \begin{minipage}{0.48\textwidth}
        \centering
        \includegraphics[width=\linewidth]{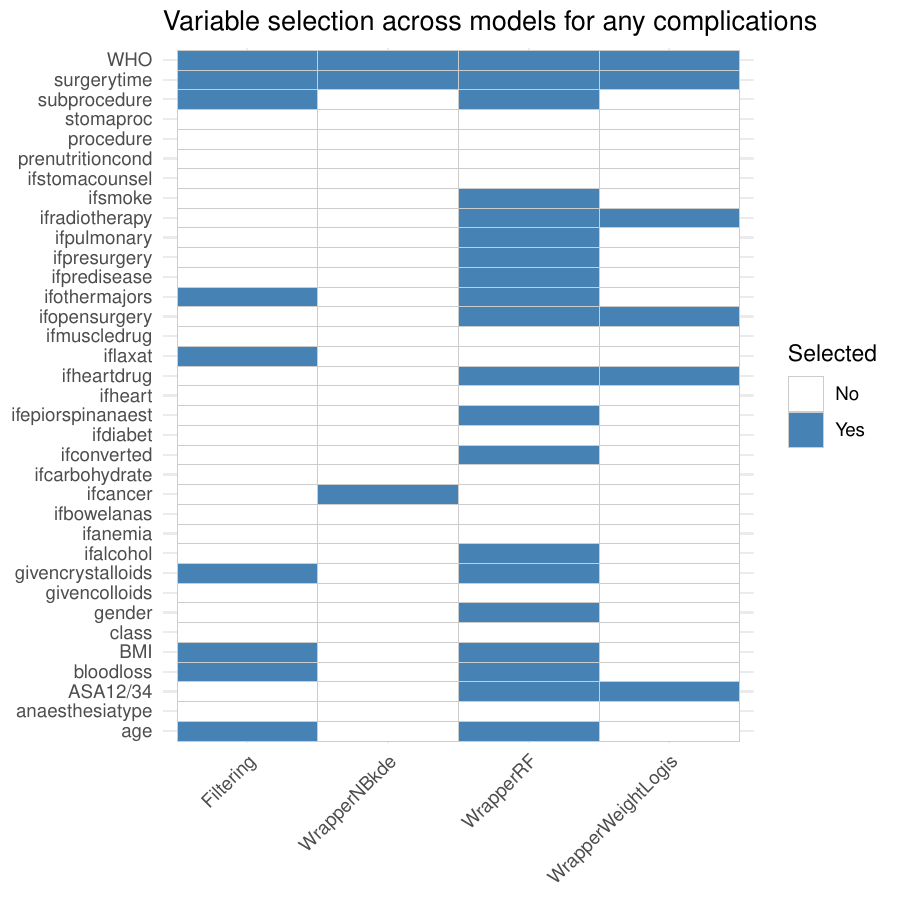} 
        \caption{Comparison of selected variables (y-axis) for predicting any complications using different variable selection methods (x-axis) on the in-sample data with 580 observations out of 34 variables.}
        \label{fig:varselcomp1}
    \end{minipage}\hfill
    \begin{minipage}{0.48\textwidth}
        \centering
        \includegraphics[width=\linewidth]{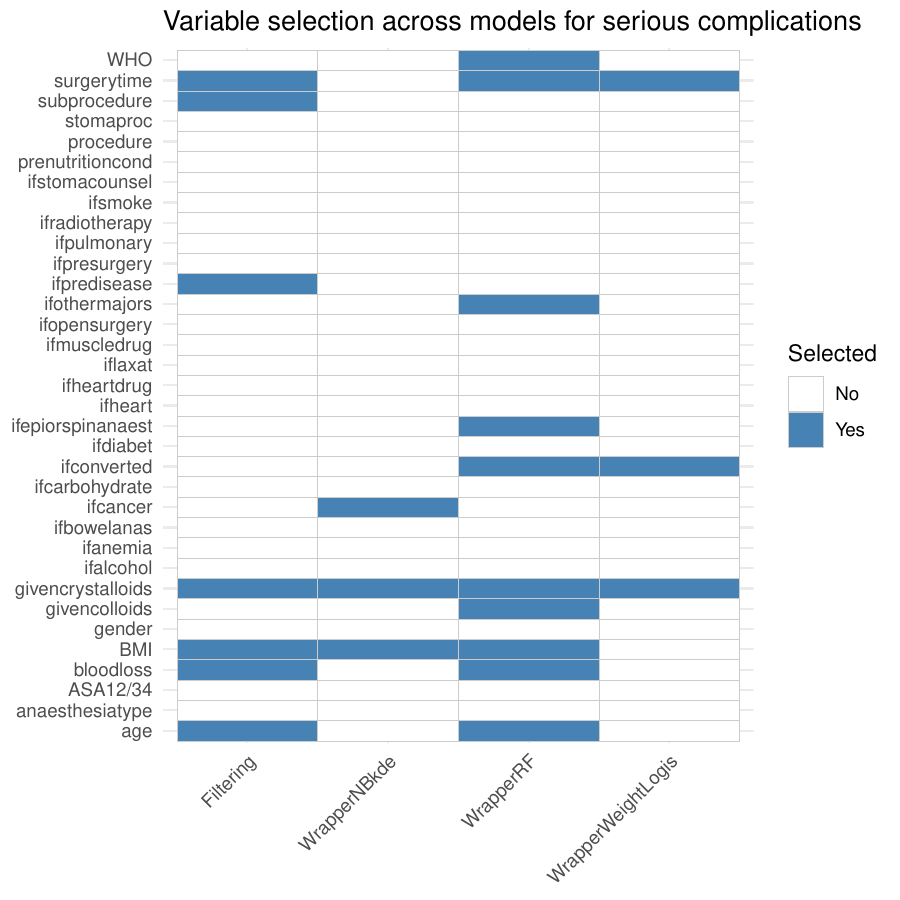} 
        \caption{Comparison of selected variables (y-axis) for predicting serious complications using different variable selection methods (x-axis) on the in-sample data with 580 observations out of 34 variables.}
        \label{fig:varselcomp2}
    \end{minipage}
\end{figure}

Mutual information (MI) is calculated between each perioperative variable and each of the considered complication variables. For \texttt{anycomp}, \texttt{WHO} exhibited the highest MI, showing the highest importance. For \texttt{seriouscomp}, \texttt{bloodloss} showed the highest MI. To capture interactions, we extended the analysis to include CMI. Conditioning on the variable with the highest MI, we select the variable with the next highest importance. This procedure can be extended to two, three, etc., conditioned variables. We continued our conditional mutual information (CMI) analysis until we conditioned on three variables. Then, we analyzed the remaining variables' CMI conditioned on the three chosen variables and chose the rest based on the elbow analysis (see Section \ref{sec:filter}) of such CMI values' plots in Appendix \ref{sec:app-CMI}. Nine variables were selected for \texttt{anycomp} and seven for \texttt{seriouscomp}, as shown in Figures \ref{fig:varselcomp1} and \ref{fig:varselcomp2}. 

The wrapper methods are based on the model employed and often apply sequential iterations. For logistic and weighted logistic models, the backward and both forward/backward led to the same selected variables regardless of the complication status and better AIC values than the forward selection in the in-sample data. Therefore, all selected variables of the logistic regression models with the wrapper analysis come from the backward direction. The selected variables are reported in Figure \ref{fig:varselcomp1} and \ref{fig:varselcomp2}. The estimated weighted logistic regression model for predicting serious complications with a wrapper-based variable selection is given in Appendix \ref{sec:app-CMI}. 
 
As detailed in Section \ref{sec:methods}, we also run the random forests with wrapper and filtering variable selection methods. Further, we provide the variable importance of random forests using all variables in Appendix \ref{sec:app-CMI}, where intraoperative variables play an important role in predictions. 


We observe in Figures \ref{fig:varselcomp1} and \ref{fig:varselcomp2}\footnote{The coefficient of the intercept was also estimated by weighted logistic regression for both outcomes, but we omit it in Figures \ref{fig:varselcomp1} and \ref{fig:varselcomp2}.} that the selected variables for the same outcome differ per method used. \texttt{WHO} and \texttt{surgerytime} are identified as risk factors for all methods to predict any complications. All methods chose more variables for predicting any complications than serious ones. If we consider the four methods and two outcomes (giving eight selections), \texttt{surgerytime} and  \texttt{givencrystalloids} were chosen as a risk factor by 75\% of the selections (six out of eight).  All three continuous variables (\texttt{surgerytime}, \texttt{BMI}, \texttt{age}) in our data were selected as risk factors by at least two of the four methods, for both outcomes.  Although discretized variables might be easier to interpret, continuous variables might better differentiate patient risks, as they contain more detailed information than binary or ordinal/nominal variables. All in all, the selected variables show the importance of both preoperative health and intraoperative events for postoperative outcomes following bowel surgeries.

\subsection{Classification performance comparison for any complications}\label{sec:predanycompresult}


The classification performance is evaluated in several dimensions, and the results are reported in Table \ref{tab:predictanycomp}, for any complications. 

For the in-sample analysis: when all selected variables are considered, the highest AUC value, of 0.79, is achieved by the Random Forests, but other AUC values are also considered acceptable \citep{de2022interpreting}. In terms of Brier score, logistic regression (LR) yields the lowest Brier score for the non-complication class. It is noteworthy, however, that this good performance of LR results from the sensitivity of the model towards the majority class, i.e., the patients exhibiting no complications. This is visible by pairing the Brier score for no complications with those for any complications. 
The two variable selection methods lead to slight changes in the performance metrics across the selected methods. The AUC depreciates in-sample with variable selection methods, with respect to the full models, and no remarkable differences in AUC are observed across the two variable selection criteria. It might be due to potential overfitting to the in-sample data. Likewise, the Brier scores of the methods slightly increase or decrease with variable selection methods across models.

For the out-of-sample setting, Naive Bayes using all variables or random forest with the wrapper variable selection is the best performer in terms of AUC.
Likewise, the wrapper method leads to an improved AUC for weighted logistic regression. All in all, we observe a modest diagnostic performance with respect to the AUC in the out-of-sample. In terms of Brier scores, (weighted) logistic regression has the best out-of-sample performance with wrapper compared to its full model and filtering-based variable selection. Even though Naive Bayes' Brier score for non-complication is better with using all variables than the wrapper, the wrapper method with Naive Bayes gives closer Brier scores for both complication and non-complication classes. Similarly, random forest using all variables has the same Brier scores for both classes.

All in all, wrapper methods give more consistent performances in the in-sample and out-of-sample for weighted logistic regression, but no major differences can be observed between the two variable selection criteria. This is notable, as the wrapper employs a selection procedure tailored for each method, whereas the filtering selection is done independently of any method. For random forest, using all variables also gives closer Brier scores across classes.

\begin{table}[ht!]
\resizebox{\textwidth}{!}{%
\begin{tabular}{l|l|lll|lll}
 \hline
 Outcome: anycomp &  & \multicolumn{3}{l|}{In-sample} & \multicolumn{3}{l}{Out-of-sample} \\ 
 \hline
 Method  & Variable set & AUC & Brier(NoComp) & Brier(AnyComp) & AUC & Brier(NoComp) & Brier(AnyComp) \\  
 \hline
 Logistic regression   & \multirow{4}{*}{All}       & 0.73 & 0.10 & 0.36 & 0.58 & 0.11 & 0.46\\
 W. logistic regression& & 0.73 & 0.20 & 0.21 & 0.58 & 0.22 & 0.30\\
 Naive Bayes (kde)     & & 0.70 & 0.19 & 0.28 & 0.64 & 0.21 & 0.32\\
 Random forests        & & 0.79 &0.20  & 0.17 &  0.63 & 0.24 & 0.24 \\  
 \hline 
  Logistic regression   & \multirow{4}{*}{Filtering}       & 0.68 & 0.10 & 0.39  & 0.58 & 0.10 & 0.47\\
 W. logistic regression & & 0.67 & 0.23 & 0.23 & 0.58 & 0.22 & 0.28\\
 Naive Bayes (kde)      & & 0.69 & 0.21 & 0.26 & 0.64 & 0.20  & 0.31 \\
 Random forests        &  & 0.78 & 0.20 & 0.18 & 0.61 & 0.23  &0.28  \\  
 \hline
   Logistic regression   & \multirow{4}{*}{Wrapper}       & 0.71 & 0.10 & 0.38  & 0.55 & 0.12 & 0.46\\
 W. logistic regression & & 0.69 & 0.22 & 0.23 & 0.63 & 0.22 & 0.26 \\
 Naive Bayes (kde)     &  & 0.67 & 0.22 & 0.23 & 0.61 & 0.23 & 0.26  \\
 Random forests        &  & 0.73 & 0.19 & 0.23 & 0.64 & 0.19 & 0.28 \\  
 \hline
\end{tabular}}
\caption{Comparison of logistic regression, weighted logistic regression, Naive Bayes with kernel density estimation, and random forests for predicting the existence of any postoperative complications using all variables or the selected variables by filtering and wrapper methods, as shown in Figure \ref{fig:varselcomp1} and \ref{fig:varselcomp2}, on in-sample (580 observations) and out-of-sample data (187 observations). The performance is measured by AUC and Brier scores for each class. Higher AUC values indicate better discriminative ability, while lower Brier scores indicate better calibration.}
\label{tab:predictanycomp}
\end{table}

\subsection{Classification performance comparison for serious complications}\label{sec:predseriouscompresult}

For modeling and predicting serious complications, the same comparative approach as in Section \ref{sec:predanycompresult} was undertaken, and the results are presented in Table \ref{tab:predictseriouscomp}. 

\begin{table}[ht!]
\resizebox{\textwidth}{!}{%
\begin{tabular}{l|l|lll|lll}
 \hline
Outcome: seriouscomp &  & \multicolumn{3}{c|}{In-sample} & \multicolumn{3}{c}{Out-of-sample} \\ 
 \hline
Method  & Variable set & AUC & Brier(NoSerComp) & Brier(SerComp) & AUC & Brier(NoSerComp) & Brier(SerComp) \\  
 \hline
  Logistic regression   & \multirow{4}{*}{All}       & 0.78 & 0.02 & 0.60 & 0.46 & 0.01 & 0.81\\
W. logistic regression &  & 0.80 & 0.18 & 0.18 & 0.54 & 0.15 & 0.47 \\
Naive Bayes (kde)      & & 0.74 & 0.18 & 0.30 & 0.67 & 0.18 & 0.40  \\
Random forests         & & 0.77 & 0.19 & 0.22 & 0.72 & 0.19 & 0.24  \\  
 \hline 
   Logistic regression   & \multirow{4}{*}{Filtering}       & 0.74 & 0.02 & 0.68  & 0.57 & 0.01 & 0.79 \\
W. logistic regression & & 0.73 & 0.22 & 0.22 & 0.59 & 0.19 & 0.32 \\
Naive Bayes (kde)      & & 0.73 & 0.18 & 0.25 & 0.70 & 0.18 & 0.30  \\
Random forests         & & 0.78 & 0.16 & 0.23 & 0.69 & 0.18 & 0.28  \\  
 \hline
    Logistic regression   & \multirow{4}{*}{Wrapper}       & 0.75 & 0.02 & 0.62 & 0.62 & 0.01 & 0.80 \\
W. logistic regression & & 0.70 & 0.21 & 0.22 & 0.68 & 0.16 & 0.30  \\
Naive Bayes (kde)      & & 0.68 & 0.21 & 0.22 & 0.60 & 0.16 & 0.33  \\
Random forests         & & 0.78 & 0.17 & 0.23 & 0.69 & 0.17 & 0.30  \\  
 \hline
\end{tabular}}
\caption{Comparison of logistic regression, weighted logistic regression, Naive Bayes with kernel density estimation, and random forests for predicting the existence of serious postoperative complications using all variables or the selected variables by filtering and wrapper methods, as shown in Figure \ref{fig:varselcomp1} and \ref{fig:varselcomp2}, on in-sample (580 observations) and out-of-sample data (187 observations). The performance is measured by AUC and Brier scores for each class. Higher AUC values indicate better discriminative ability, while lower Brier scores indicate better calibration.}
\label{tab:predictseriouscomp}
\end{table}

For the in-sample analysis, in terms of AUC, a good discriminative performance of 0.80 is obtained by the weighted logistic regression with the full model. Employing variable selection methods leads to a drop in AUC values for all the models, except for Random Forests. Even more exacerbated than for any complications, logistic regression shows very small Brier scores for predicting non-serious complications. This is, again, due to the strong bias towards non-serious complications. Using variable selection methods leads to worse Brier scores for weighted logistic regression and mixed results for Naive Bayes and random forests.


For the out-of-sample analysis, the full models show a deteriorating discriminative performance compared to the in-sample, with only Random Forests showing an AUC value higher than 0.70. The AUC value of (weighted) logistic regressions improved with both filtering and wrapper compared to its full model. For Naive Bayes, out-of-sample, the filtering leads to a better AUC, whereas the wrapper leads to a lower AUC. For Random Forests, the best out-of-sample AUC is achieved for the full model. Regarding the Brier scores, Random Forests using all variables have closer Brier scores for both classes.


Overall, for weighted logistic regression and Naive Bayes, the wrapper and filtering lead to the best out-of-sample predictive performance in predicting serious complications, respectively. Still, regarding the similar performances in the in-sample and out-of-sample, Random Forest using all variables is the best among the considered models and variable selection methods.

\section{Discussion and conclusion}\label{sec:conc}
In this study, we compared three probabilistic classifiers, weighted logistic regression, random forests, and Naive Bayes, for predicting both any and serious postoperative complications following elective bowel surgery. In our analyses, we accounted for class imbalance and evaluated the role of variable selection methods.

Random forests yielded the best predictive performance across both outcomes when all 34 perioperative variables were considered. Weighted logistic regression performed better than its unweighted counterpart under class imbalance, while Naive Bayes offered comparable performance. The impact of variable selection was method-dependent: wrapper approaches (e.g., stepwise selection) provided modest improvements for weighted logistic regression but did not substantially enhance the performance of random forests.  Still, it should be emphasized that a preselection of the variables, leading to the 34 considered variables, has been employed, and its influence on the overall performance of the models should not be overlooked. 

Even though both outcomes shared common variables, such as surgery duration, age, and blood loss, we also observed some differences: for instance, the amount of colloids administration was associated with serious complications but not with the broader any complication outcome.

To assess model performance probabilistically, we used class-specific Brier scores rather than threshold-based classification accuracy. It allows for direct quantification of predictive uncertainty. We advocate reporting such probabilities in clinical decision-making dashboards: patients with confidently low predicted risk (e.g., lower than 0.15) might be eligible for accelerated discharge, whereas those with high or uncertain risk could be flagged for enhanced monitoring. In scenarios where predictive uncertainty remains high, data-driven results could be complemented with additional knowledge, such as expert judgment. There is a diversity of methods that elicit and aggregate expert judgments. Of these, the Classical Model for Structured Expert Judgment \citep{cooke1991experts} proposes a mathematical model to aggregate expert uncertainty estimates, which employs a validated weighting scheme. By employing the Classical Model, it can be ensured that experts' assessments are subjected to the same empirical validation as the data-driven methods. 

Our study has limitations. It is based on data from a single center, and the number of serious complications was relatively small. External validation in larger, multi-center cohorts is essential. Future research should also explore alternative interpretable modeling frameworks, such as generalized additive models or Bayesian networks; investigate the clinical impact and cost-effectiveness of integrating probabilistic forecasts into routine care; and develop tools that transparently communicate predictive uncertainty to clinicians and patients. Moreover, modeling first-order interactions in (weighted) logistic regression can be analyzed, and although protocols did not change, inclusion of the COVID-19 period may have introduced temporal shifts that we could not model given our limited sample size.

\subsection*{Acknowledgement}\label{sec:ack}
We thank Reini Bretveld for the data access used in this study. We are also grateful to Ilse Waanders for her input on the clinical relevance. The authors used Grammarly for language editing, but all scientific content is entirely their own.

\subsection*{Statements and declarations}
\textbf{Declaration of conflicting interest}: All authors have none to declare.

\textbf{Funding statement}:  This work is part of the 4TU programme RECENTRE (Risk-based lifEstyle Change: daily-lifE moNiToring and REcommendations). It is funded by the 4TU programme High Tech for a Sustainable Future. 4TU is the federation of the four technical universities in the Netherlands (Delft University of Technology, Eindhoven University of Technology, University of Twente, Wageningen University and Research). 

\textbf{Ethical considerations}: This study received ethical approval from the niet-WMO adviescommissie MST  (approval K24-41) on Month 12, 2024. This is an IRB-approved retrospective study, all patient information was de-identified and patient consent was not required. Patient data will not be shared with third parties.

\textbf{Consent to participate and for publication}: Informed consent was not required for this retrospective study. The study uses pseudonymized data collected between 2009 and 2023 from the ERAS registry. Obtaining individual consent would be unreasonably difficult and, in many cases, impossible due to the long study period, the large number of patients, and the fact that some individuals may have passed away or can no longer be contacted. The data are fully pseudonymized and were collected as part of routine clinical care, with no additional burden placed on patients. This study qualifies as non-WMO research and complies with ethical standards for the use of retrospective, non-identifiable data.

\textbf{Data availability}: The data that support the findings of this study are not publicly available due to confidentiality agreements with the data provider, but may be obtained from the corresponding author upon reasonable request.

\appendix

\section{Further data preprocessing steps}\label{sec:app-datapreproc}

\textbf{Variable encoding steps}

\begin{itemize}
    \item Patients were classified as diabetic if they managed their condition through either diet or medication.
    \item  For patients with documented termination of smoking, a history of smoking was assumed.
    \item Alcohol consumption was assumed for patients with documented termination.
    \item Due to low representation (approximately 2\%), ASA classes 1 and 4 were merged with classes 2 and 3.
    \item Nutritional status was encoded as binary to indicate malnutrition; risk of malnutrition is included as malnutrition.
    \item Carbohydrate intake was encoded as a binary variable (yes/no) independent of the intake method.
    \item Planned surgery types were encoded as either open or laparoscopic\&robotic\&stoma.
    \item Blood loss was categorized into three groups: "No loss," "$\leq 100$ mL," and "$>100$ mL".
    \item  Intraoperative crystalloid volume was also categorized into three groups:  "$< 1000$ mL," "$ \geq1000 \text{ and } < 2000$ mL," and "$\geq2000$ mL".
    \item Stoma-related procedures were coded into 0 (none or unrelated) and 1 (colostomy- or ileostomy-related).
    \item Major concurrent procedures (\texttt{ifothermajors}) were encoded as binary (yes/no).
    \item Colloid administration (\texttt{givencolloid})  was similarly encoded as 0 and 1, as only 17\% of patients received colloids.
    \item  Subprocedures (\texttt{subprocedure}) were merged under four categories called rectosigmoid (abdominoperineal resection, anterior resection of rectum, proctocolectomy with anus, sigmoid resection), stoma procedure and stoma reversal (Other stoma procedures, reversal of Hartmann's procedure),  other (small bowel resection, other large/small bowel surgery, exploratory laparotomy/laparoscopy only, other), and partial colectomy (ileocaecal/right hemicolectomy,  left hemicolectomy, total/subtotal colectomy). 
    \item 4 unknown epidural or spinal anesthesia (\texttt{ifepiorspinanaest}), and 12 muscle relaxant (\texttt{ifmuscledrug}) and heart drug (\texttt{ifheartdrug}) usage values are taken as "No."
\end{itemize}

\begin{table}[H]
\centering
\begin{tabular}{|l|l|l|}
\hline
\textbf{Imputed var.} & \textbf{\# Missing} & \textbf{Conditional table} \\ \hline
\texttt{BMI} & 63 & Discretized \texttt{age}, \texttt{gender}, \texttt{ifdiabet} \\ \hline
\texttt{ifsmoke} & 4 & Discretized \texttt{age}, discretized \texttt{BMI}, \texttt{gender} \\ \hline
\texttt{ifalcohol} & 7 & Discretized \texttt{age}, discretized \texttt{BMI}, \texttt{gender} \\ \hline
\texttt{ASA} & 6 & Discretized \texttt{age}, discretized \texttt{BMI} \\ \hline
\texttt{WHO} & 44 & Discretized \texttt{age}, discretized \texttt{BMI}, \texttt{gender} \\ \hline
\texttt{prenutritioncond} & 48 & Discretized \texttt{age}, discretized \texttt{BMI}, \texttt{ASA}, \texttt{gender} \\ \hline
\texttt{ifpresurgery} & 3 & Discretized \texttt{age}, discretized \texttt{BMI}, \texttt{ASA}, \texttt{ifpredisease} \\ \hline
\texttt{ifstomacounsel} & 1 & Discretized \texttt{age}, discretized \texttt{BMI}, \texttt{ASA}, \texttt{ifpredisease} \\ \hline
\texttt{ifcarbohydrate} & 1 & Discretized \texttt{age}, discretized \texttt{BMI}, \texttt{ASA}, \texttt{ifpredisease} \\ \hline
\texttt{iflaxat} & 6 & Discretized \texttt{age}, discretized \texttt{BMI}, \texttt{ASA}, \texttt{ifpredisease} \\ \hline
\texttt{ifanemia} & 79 & Discretized \texttt{age}, discretized \texttt{BMI}, \texttt{ASA}, \texttt{finaldiagnosis} \\ \hline
\texttt{bloodloss} & 4 & Discretized \texttt{surgerytime}, \texttt{procedure} \\ \hline
\texttt{ifothermajors} & 2 &  \texttt{bloodloss}, discretized  \texttt{surgerytime}, \texttt{procedure} \\ \hline
\texttt{givencrystalloids} & 16 & \texttt{bloodloss}, discretized \texttt{surgerytime}, \texttt{procedure} \\ \hline
\texttt{givencolloids} & 16 & \texttt{bloodloss}, discretized \texttt{surgerytime}, \texttt{procedure} \\ \hline
\texttt{anaesthesiatype} & 8 & \texttt{ifnerveorlocalanaest}, \texttt{ifepiorspinanaest}, \texttt{procedure} \\ \hline
\end{tabular}
\caption{Imputations based on conditional tables for missing data in variables. The table lists the imputed variables, the number of missing observations, and the conditional variables used for imputation. The discretization of continuous variables includes  \texttt{surgerytime} grouped into "Short" (less than 1.5 hours), "Medium" (between 1.5 and 2.5 hours), and "Long" (more than 2.5 hours); \texttt{age} grouped into 10-year intervals (e.g., "20-29", "30-39", etc.); and \texttt{BMI} grouped into four categories such as "Underweight" (until 18.5), "Normal weight" (18.5 - 24.9), "Overweight" (25.0 - 29.9), and "Obese" (more than 29.9). We use the estimated conditional mean for continuous variables, whereas the conditional mode is for discrete ones.}
\label{tab:app-imputation}
\end{table}

\section{Descriptive summaries}\label{sec:app-datasummary}

\begin{table}[H]
\centering
\tiny
\scalebox{0.95}{%
\begin{tabular}{lllll}
\hline
\textbf{Variable} & 
  \textbf{$j=0$} & 
  \textbf{$j=1$ } &
  \textbf{$j=0_s$} &
  \textbf{$j=1_s$}  \\
\hline
\texttt{age}                & 65.43 [24--95]      & 64.92 [20--82]    & 64.90  [20--95]   & 68.21 [34--86]\\ \hline
\texttt{ifalcohol}          & 57 (49\%)  & 32 (45\%) & 77 (46\%) & 12 (63\%) \\ \hline
\texttt{BMI}                & 26.44   [16.39--37.54]   & 26.07   [16.14--39.18]  & 26.16  [16.14--39.18]   & 27.55 [20.42--35.06]\\ \hline
\texttt{ASA12/34}           & \begin{tabular}[l]{@{}l@{}}
1: 72 (62\%) \\ 3: 44 (38\%)
 \end{tabular}       & 
\begin{tabular}[l]{@{}l@{}}
1: 47 (66\%) \\ 3: 24 (34\%)
 \end{tabular}   & \begin{tabular}[l]{@{}l@{}}
1: 108 (64.3\%) \\ 3: 0 (35.7\%)
 \end{tabular} & 
 \begin{tabular}[l]{@{}l@{}}
 1: 11 (57.9\%) \\ 3: 8 (42.1\%)
 \end{tabular} \\ \hline
\texttt{gender}             & 53 (46\%)  & 30 (42\%) & 77 (46\%) & 6 (31.6\%) \\ \hline
\texttt{ifpredisease}       & 32 (28\%)  & 19 (27\%) & 43 (27.6\%) & 8 (42.1\%) \\ \hline
\texttt{ifdiabet}           & 17 (15\%)  & 10 (14\%) & 22 (13\%) & 5 (26.3\%) \\ \hline
\texttt{ifheart}            & 12 (10\%)  & 9 (13\%)  & 18 (10.7\%) & 3 (15.8\%) \\ \hline
\texttt{ifpulmonary}        & 9 (8\%)    & 3 (4\%)   & 10 (6\%)   & 2 (10.5\%) \\ \hline
\texttt{WHO}                & 5 (4\%)    & 2 (3\%)   & 7 (4\%) & 0 (0\%) \\ \hline
\texttt{ifsmoke}            & 26 (22\%)  & 16 (23\%) & 39 (23.2\%) & 3 (15.8\%) \\ \hline
\texttt{prenutritioncond}   & 23 (20\%)  & 12 (17\%) & 32 (19\%) & 3 (15.8\%) \\ \hline
\texttt{ifpresurgery}       & 51 (44\%)  & 32 (45\%) & 69 (41\%) & 14 (73.7\%) \\ \hline
\texttt{ifradiotherapy}     & 7 (6\%)    & 4 (6\%)   & 8 (4.8\%)  & 3 (15.8\%) \\ \hline
\texttt{iflaxat}            & 40 (34\%)  & 4 (6\%)   & 30 (17.9\%)  & 0 (0\%) \\ \hline
\texttt{ifstomacounsel}     &   40 (34.5\%)         &      24 (33.8\%)     &   57 (34\%)        & 7 (36.7\%) \\ \hline
\texttt{ifcarbohydrate}     & 99 (85\%)  & 62 (87\%) & 146 (87\%) & 15 (79\%) \\ \hline
\texttt{ifanemia}           & 36 (31\%)  & 30 (42\%) & 59 (35.1\%) & 7 (36.8\%) \\ \hline
\texttt{ifopensurgery}      & 20 (17\%)  & 47 (66\%) & 35 (21\%) & 8 (42.1\%) \\ \hline
\texttt{ifothermajors}      & 26 (22\%)  & 23 (32\%) & 41 (24.4\%) & 8 (42.1\%) \\ \hline
\texttt{ifbowelanas}        & 92 (79\%)  & 47 (66\%) & 126 (75\%)  & 13 (68.4\%) \\ \hline
\texttt{ifmuscledrug}       & 41 (35\%)  & 24 (34\%) & 57 (34\%) & 8 (42\%) \\ \hline
\texttt{anaesthesiatype}    &
\begin{tabular}[l]{@{}l@{}}
1: 8 (6.9\%)\\ 2: 108 (93.1\%)
 \end{tabular}

 & \begin{tabular}[l]{@{}l@{}}1: 7 (9.9\%)\\ 2: 64 (90.1\%) \end{tabular}& \begin{tabular}[l]{@{}l@{}}1: 13 (7.7\%)\\ 2: 155 (92.3\%) \end{tabular} & \begin{tabular}[l]{@{}l@{}}1: 2 (10.5\%)\\ 2: 17 (89.5\%)\end{tabular} \\ \hline
\texttt{ifheartdrug}        & 104 (90\%) & 66 (93\%) & 153 (91\%) & 17 (89.5\%) \\ \hline
\texttt{bloodloss}          & \begin{tabular}[l]{@{}l@{}}1: 38 (54\%) \\ 2: 12 (17\%) \\ 3: 21 (30\%)  \end{tabular}        &  \begin{tabular}[l]{@{}l@{}}1: 83 (71.5\%) \\ 2: 15 (13\%) \\ 3: 18 (15.5\%)  \end{tabular}&  \begin{tabular}[l]{@{}l@{}}1: 112 (66.7\%) \\ 2: 24 (14.3\%) \\ 3: 32 (19.0\%)  \end{tabular}&  \begin{tabular}[l]{@{}l@{}}1: 9 (47.3\%)\\ 2: 3 (15.8\%)\\ 3: 7 (36.9\%) \end{tabular}\\ \hline
\texttt{ifgivencolloids}      & 16 (13.8\%)        &   8 (11.3\%)        & 21 (12.5\%) & 3 (15.8\%)\\ \hline
\texttt{givencrystalloids}  & \begin{tabular}[l]{@{}l@{}}1: 93 (80\%) \\ 2: 21 (18\%) \\ 3: 2 (2\%) \end{tabular}        &  \begin{tabular}[l]{@{}l@{}}1: 50 (70\%) \\ 2: 17 (24\%) \\ 3: 4 (6\%) \end{tabular}&  \begin{tabular}[l]{@{}l@{}}1: 130 (77.4\%) \\ 2: 35 (20.8\%) \\ 3: 3 (1.8\%)\end{tabular} &  \begin{tabular}[l]{@{}l@{}}1: 13 (68.4\%)\\ 2: 3 (15.8\%) \\ 3: 3 (15.8\%) \end{tabular}\\ \hline
\texttt{surgerytime}        & 121.68 [21--324]     & 148.1  [21--327]  & 127.70 [21--327]    & 167.4 [21--290]\\ \hline
\texttt{procedure}          & \begin{tabular}[l]{@{}l@{}}1: 104 (90\%) \\ 2: 12 (10\%) \end{tabular}        &  \begin{tabular}[l]{@{}l@{}}1: 64 (90\%) \\ 2: 7 (10\%) \end{tabular} &  \begin{tabular}[l]{@{}l@{}}1: 151 (90\%) \\ 2: 17 (10\%) \end{tabular} &  \begin{tabular}[l]{@{}l@{}}1: 17 (89.4\%)\\ 2: 2 (10.6\%) \end{tabular}\\ \hline
\texttt{ifconverted}        & 0  (0\%)     &   1 (1.4\%)        & 1  (0.6\%)       & 0 (0\%) \\ \hline
\texttt{stomaproc}          & 38 (33\%)  &     26 (36.7\%)      & 58 (34.5\%) & 6 (31.6\%) \\ \hline
\texttt{ifcancer}           & 82 (71\%)  &     45 (63.4\%)      & 117 (69.6\%) & 10 (52.6\%) \\ \hline

\texttt{ifepiorspinanaest}
 & \begin{tabular}[c]{@{}l@{}}
    0: 22 (19\%)\\
    1: 86 (74\%)\\
    2: 8 (7\%)
   \end{tabular}
 & \begin{tabular}[c]{@{}l@{}}
    0: 8 (11\%)\\
    1: 49 (69\%)\\
    2: 14 (20\%)
   \end{tabular}
 & \begin{tabular}[c]{@{}l@{}}
    0: 27 (16\%)\\
    1: 124 (74\%)\\
    2: 17 (10\%)
   \end{tabular}
 & \begin{tabular}[c]{@{}l@{}}
    0: 3 (16\%)\\
    1: 11 (58\%)\\
    2: 5 (26\%)
   \end{tabular} \\
 \hline

\texttt{subprocedure}       & \begin{tabular}[l]{@{}l@{}}1: 46 (40\%)\\ 2: 23 (20\%)\\ 3: 7 (6\%)\\ 4: 40 (34\%) \end{tabular}        & \begin{tabular}[l]{@{}l@{}}1: 18 (25\%)\\ 2: 13 (18\%)\\ 3: 12 (17\%)\\ 4: 28 (39\%) \end{tabular}& \begin{tabular}[l]{@{}l@{}}1: 59 (35.1\%)\\ 2: 34 (20.2\%)\\ 3: 16 (9.5\%)\\ 4: 59 (35.1\%) \end{tabular}& \begin{tabular}[l]{@{}l@{}}1: 5 (26.3\%)\\ 2: 2 (10.6\%)\\ 3: 3 (15.8\%)\\ 4: 9 (47.3\%)\end{tabular}\\ \hline
\end{tabular}
}
\caption{Summary of variables in not developing (116 observations, $j=0$) and developing any complications (71 observations, $j=1$), as well as in not developing (168 observations, $j=0_s$) and developing serious complication classes (19 observations, $j=1_s$) in out-of-sample data. For each variable, it provides the distribution or summary measure (range and mean) based on the variable type, along with a brief description of the variable. For binary variables taking values of 0 and 1, the summaries for the value of 1 are reported.}
\label{tab:summaryout}
\end{table}

\section{Variable selection and model fit results}\label{sec:app-CMI}
\begin{figure}[H]
    \centering
    \includegraphics[width=.5\linewidth]{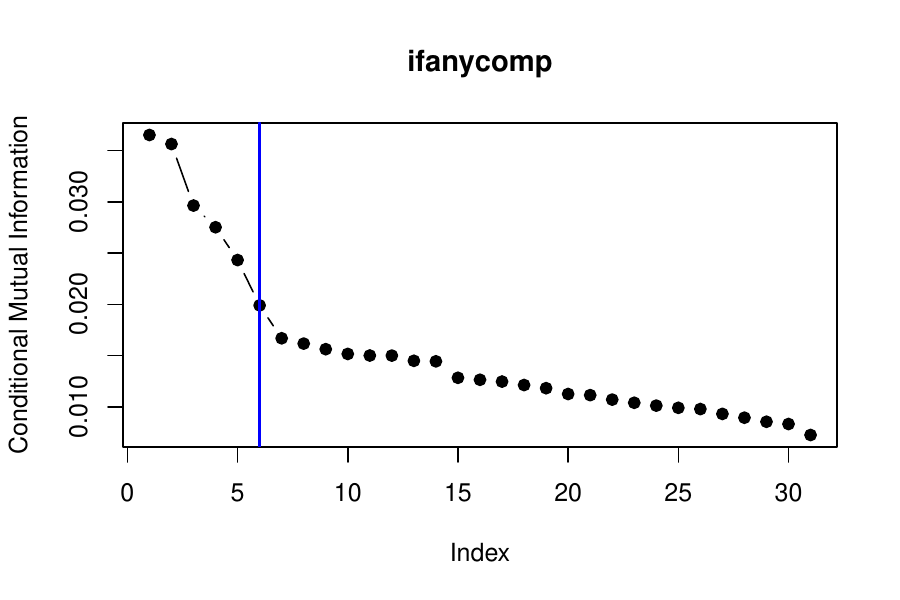} 
        \includegraphics[width=.5\linewidth]{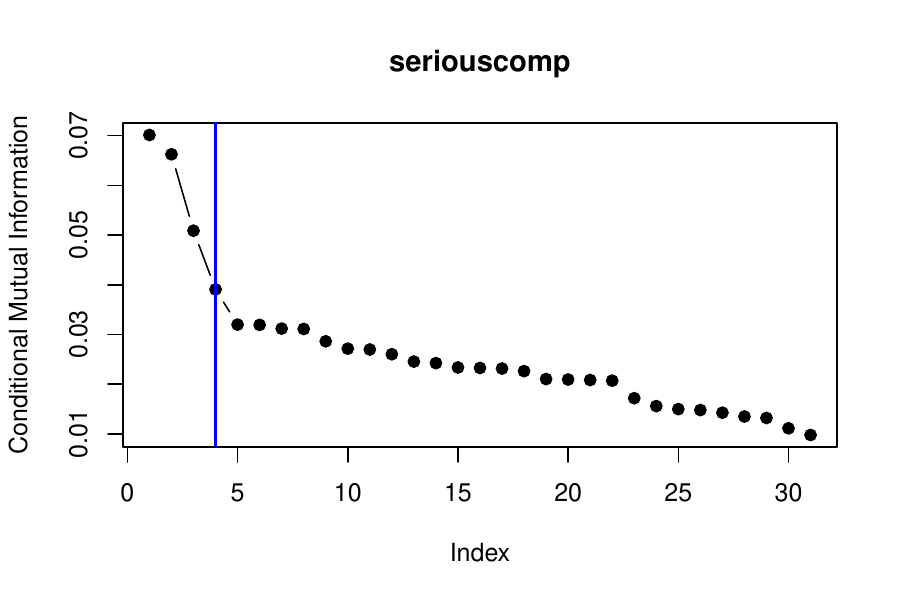}

    \caption{Conditional mutual information analysis of remaining variables for the outcome of any complications (top) and serious complications (bottom). The CMI values for each variable, conditioned on three previously selected variables, to identify additional influential variables, are given on the y-axis. The elbow index, indicated by a vertical line, highlights the threshold we used to select significant variables for further analysis, including that index's variable.}
    \label{fig:cmi_ifanycomp}
\end{figure}


Selected variables by CMI are as follows in their selection order:

For \texttt{anycomp}:
\texttt{WHO},
  \texttt{ifothermajors},
  \texttt{surgerytime},
\texttt{subprocedure},
  \texttt{BMI},
    \texttt{age},
    \texttt{bloodloss},
    \texttt{givencrystalloids}, and
\texttt{iflaxat}.

For \texttt{seriouscomp}:
 \texttt{bloodloss},
     \texttt{surgerytime},
    \texttt{subprocedure},
    \texttt{BMI},
    \texttt{age},
    \texttt{givencrystalloids}, and
    \ \texttt{ifpredisease}.

\begin{table}[H]
\centering
\begin{minipage}[t]{0.48\linewidth}
\centering
\scalebox{0.8}{
\begin{tabular}{lr}
\hline
\multicolumn{2}{c}{AnyComp} \\
\hline
Variable& MDG \\
\hline
\texttt{surgerytime}       & 6.46 \\
\texttt{age}               & 4.06 \\
\texttt{BMI}               & 4.04 \\
\texttt{WHO}               & 2.71 \\
\texttt{givencrystalloids} & 2.51 \\
\texttt{ifothermajors}     & 2.23 \\
\texttt{ASA12.34}          & 1.19 \\
\texttt{ifepiorspinanaest} & 1.11 \\
\texttt{bloodloss}         & 1.11 \\
\texttt{ifopensurgery}     & 0.83 \\
\texttt{ifpulmonary}       & 0.78 \\
\texttt{ifradiotherapy}    & 0.71 \\
\texttt{ifsmoke}           & 0.60 \\
\texttt{subprocedure}      & 0.59 \\
\texttt{ifalcohol}         & 0.58 \\
\texttt{ifpredisease}      & 0.48 \\
\texttt{anaesthesiatype}   & 0.48 \\
\texttt{ifheartdrug}       & 0.47 \\
\texttt{ifpresurgery}      & 0.46 \\
\texttt{ifconverted}       & 0.46 \\
\texttt{iflaxat}           & 0.41 \\
\texttt{ifstomacounsel}    & 0.40 \\
\texttt{ifgivencolloids}     & 0.39 \\
\texttt{gender}            & 0.35 \\
\texttt{ifbowelanas}       & 0.35 \\
\texttt{ifcancer}          & 0.35 \\
\texttt{prenutritioncond}  & 0.33 \\
\texttt{ifanemia}          & 0.29 \\
\texttt{stomaproc}         & 0.27 \\
\texttt{ifheart}           & 0.26 \\
\texttt{ifmuscledrug}      & 0.25 \\
\texttt{ifcarbohydrate}    & 0.22 \\
\texttt{ifdiabet}          & 0.18 \\
\texttt{procedure}         & 0.18 \\
\hline
\end{tabular}}
\end{minipage}%
\hfill
\begin{minipage}[t]{0.48\linewidth}
\centering
\scalebox{0.8}{
\begin{tabular}{lr}
\hline
\multicolumn{2}{c}{SerComp} \\
\hline
Variable      &MDG \\

\hline
\texttt{surgerytime}       & 1.73 \\
\texttt{bloodloss}         & 1.31 \\
\texttt{BMI}               & 1.19 \\
\texttt{givencrystalloids} & 1.18 \\
\texttt{age}               & 0.89 \\
\texttt{ifothermajors}     & 0.55 \\
\texttt{ifconverted}       & 0.50 \\
\texttt{ifepiorspinanaest} & 0.35 \\
\texttt{ifgivencolloids}     & 0.30 \\
\texttt{WHO}               & 0.25 \\
\texttt{ifstomacounsel}    & 0.23 \\
\texttt{ifopensurgery}     & 0.23 \\
\texttt{ifpresurgery}      & 0.20 \\
\texttt{ifdiabet}          & 0.19 \\
\texttt{ifanemia}          & 0.19 \\
\texttt{subprocedure}      & 0.18 \\
\texttt{procedure}         & 0.17 \\
\texttt{ifmuscledrug}      & 0.17 \\
\texttt{ifcancer}          & 0.16 \\
\texttt{ifpulmonary}       & 0.13 \\
\texttt{ifsmoke}           & 0.12 \\
\texttt{ASA12.34}          & 0.11 \\
\texttt{ifheartdrug}       & 0.11 \\
\texttt{iflaxat}           & 0.10 \\
\texttt{ifpredisease}      & 0.09 \\
\texttt{ifalcohol}         & 0.09 \\
\texttt{gender}            & 0.07 \\
\texttt{prenutritioncond}  & 0.07 \\
\texttt{ifcarbohydrate}    & 0.07 \\
\texttt{stomaproc}         & 0.07 \\
\texttt{ifheart}           & 0.06 \\
\texttt{ifradiotherapy}    & 0.05 \\
\texttt{ifbowelanas}       & 0.05 \\
\texttt{anaesthesiatype}   & 0.05 \\
\hline
\end{tabular}}
\end{minipage}
\caption{Random forest variable importance given by the mean decrease in Gini index (MDG) for predicting any and serious complications in the in-sample data (580 observations; 34 variables).}
\label{tab:rf_import_combined}
\end{table}

In the weighted logistic regression fit for predicting serious complications with backward variable selection: the intercept is estimated at $-1.1631$, with increased odds of serious complications associated with higher volumes of crystalloids (coefficients of $0.2868$ and $1.3500$ for levels 2 and 3, respectively), longer surgery time (coefficient $0.0049$), and surgical conversion (coefficient $1.1845$). The model fit yields a null deviance of 153.5 on 579 degrees of freedom, a residual deviance of 137.7 on 575 degrees of freedom, and an AIC of 88.03.

\raggedright

\bibliography{medrisk}
\end{document}